\documentclass[twocolumn,apl,amsmath,amssymb,showpacs,superscriptaddress]{revtex4-2}
\usepackage{epsf}      
\usepackage{graphicx}
\usepackage{color}
\usepackage{soul}
\usepackage{gensymb}
\usepackage{sidecap}
\usepackage{amsmath}
\usepackage{mathtools}
\usepackage{multirow}
\usepackage[hidelinks,colorlinks=true,linkcolor=blue,citecolor=blue]{hyperref}

\begin{document}
\title{Order-disorder transition and Na-ion redistribution in NASICON-type Na$_3$FeCr(PO$_4$)$_3$}
\author{Madhav Sharma}
\affiliation{Department of Physics, Indian Institute of Technology Delhi, Hauz Khas, New Delhi-110016, India}
\author{Archna Sagdeo}
\affiliation{Indus Synchrotrons Utilization Division, Raja Ramanna Centre for Advanced Technology, Indore-452013, India}
\author{Rajendra S. Dhaka}
\email{rsdhaka@physics.iitd.ac.in}
\affiliation{Department of Physics, Indian Institute of Technology Delhi, Hauz Khas, New Delhi-110016, India}

\date{\today}
\begin{abstract}

We report the temperature-dependent synchrotron based X-ray diffraction analysis of NASICON type Na$_3$FeCr(PO$_4$)$_3$ sample, which undergoes a symmetry-lowering structural transition from a monoclinic ($C2/c$) phase with long-range Na-vacancy order to a rhombohedral ($R\bar{3}c$) phase with statistical disordered Na ions. The [FeCr(PO$_4$)$_3$] polyanionic framework remains essentially unchanged, confirming that the transition is governed by redistribution of the Na sublattice rather than by reconstruction of the host framework. The structural evolution is accompanied by a discontinuous increase in the $c$-axis and the unit-cell volume, reflecting the progressive depopulation of the Na(1) sites and transfer of Na ions toward the Na(2) sublattice. The temperature dependence of superstructure intensity found to deviate from  mean-field critical behavior, instead, the experimental evolution is accurately captured by a sigmoidal phase-fraction model. 
The calorimetric measurements show that the enthalpy change for the first transition around 350~K is significantly larger than that of the anomaly around 445 K, indicating  the dominant configurational rearrangement of Na ions occurs within the lower-temperature interval. Overall, the diffraction and calorimetric results demonstrate that Na ordering proceeds through an order-disorder transition involving intermediate Na configurations and a finite coexisting regime. The quantitative correlation between Na-vacancy ordering, lattice strain, and symmetry lowering reveals the central role of configurational interactions within the Na conduction channels in governing the phase stability of NASICON-type materials.

\end{abstract}
\maketitle
\section{\noindent~Introduction}

Sodium-ion batteries have recently gained considerable attention as a viable alternative to lithium-based systems, sparking broad interest in how their cathode materials behave at a fundamental level \cite{Jin_CSR_20, Sharma_CCR_25}. The electrochemical performance of these materials is governed by the interplay of several key factors, such as crystal structure, electronic configuration, the chemical environment of transition metals, and the positions of mobile sodium ions (Na$^+$) within the lattice \cite{Liu_MT_16}. In certain materials, this interplay produces remarkable ordering phenomena, including Na$^+$-vacancy, transition-metal (e.g., honeycomb or cation), orbital, and charge ordering, which may be intrinsic or emerge gradually during electrochemical operation \cite{Park_CM_21, Yang_NRC_26, Yan_NC_19}. Such ordering often leads to the formation of superstructures that encode site preferences and lattice interactions beyond the average crystal structure \cite{Nishimura_ACIE_15, Boucher_JACS_14}. These ordering phenomena frequently coexist and combine to produce complex structural features. Among them, the spatial distribution of Na ions and the vacancies are especially consequential, as they directly control how ions move through the material and how stable the structure remains over time \cite{Redhammer_ACB_21}.

It has also been reported that coupling between Na$^+$ ordering and the transition-metal sublattice contributes to structural stability during battery cycling \cite{Wang_ACIE_24}. On the other hand, several reports have highlighted that excessively rigid ordering can impede ion transport, a finding that has motivated strategies such as aliovalent or isovalent substitution to regulate long-range ordering and improve diffusion kinetics \cite{Wang_SA_18}. Moreover, it has been established that the arrangement of Na ions around transition-metal centers can modify metal-oxygen covalency through a secondary inductive effect, thereby influencing the redox potential and working voltage of cathode materials \cite{Melot_ACSAMI_14}. Also, studies have further shown that an optimal balance between ordering and disordering facilitates percolation pathways for Na$^+$ migration while tuning the local electronic environment of transition-metal ions in a way that enhances operating voltages \cite{Shi_NC_22}. Building on these considerations, the literature has demonstrated that controlled Na$^+$-vacancy ordering strongly influences ionic transport, redox energetics, and structural stability, with direct consequences for both rate capability and energy density in batteries \cite{He_AEM_25}. These observations establish Na$^+$ ordering as a key parameter for tailoring the electrochemical performance of cathode materials in sodium-ion batteries.

In this context, NASICON (NAtrium Super Ionic CONductor)-type materials provide an ideal platform to explore the coupling between ionic ordering and lattice symmetry. These materials are among the most extensively studied cathodes and solid electrolytes for sodium-ion batteries due to their robust crystal structure, open framework, and excellent electrochemical stability \cite{Sharma_CCR_25, Sapra_WEE_21}. With a general formula A$_x$MM$'$(XO$_4$)$_3$ (where A is an alkali metal, M is a transition metal, and X is P, Si, or S), the NASICON compounds consist of a rigid three-dimensional framework formed by corner-sharing XO$_4$ tetrahedra and MO$_6$ octahedra. This framework creates two distinct Na sites, Na(1) and Na(2), that enable three-dimensional diffusion pathways. The sixfold-coordinated Na(1) site is located between adjacent octahedral units along the $c$ axis, while the eightfold-coordinated Na(2) site resides along an $a$-axis-like diffusion channel between neighboring Na(1) sites \cite{Meena_S_25, Sharma_S_25}. Goodenough and Hong demonstrated that NASICON compounds like Na$_{1+x}$Zr$_2$Si$_x$P$_{3-x}$O$_{12}$ exhibit a strong dependence of crystal symmetry on Na content, with monoclinic structures stabilized near three Na per formula unit and rhombohedral symmetry favored at lower or higher Na concentrations \cite{Hong_MRB_76}. Consistently, the Na$^+$-vacancy ordering has been widely reported in NASICON compounds with three sodium ions per formula unit, including Na$_3$Ti$_2$(PO$_4$)$_3$, Na$_3$Fe$_2$(PO$_4$)$_3$, Na$_3$V$_2$(PO$_4$)$_3$, and Na$_3$FeV(PO$_4$)$_3$ \cite{Kabbour_JACS_11, d’Yvoire_SSI_83, Chotard_CM_15, Park_CM_21}, leading to monoclinic distortion ($\alpha$-NASICON) at low temperatures and stabilization of the rhombohedral phase ($\gamma$-NASICON) at elevated temperatures. Therefore, investigating the temperature-dependent evolution of such ordered states provides direct insight into Na-ion configurations and their role in governing structural and transport properties. 

Understanding how the mobile ions reorganize within a rigid NASICON-type polyanionic framework is vital for controlling ion transport in battery materials for energy storage applications. Therefore, in this paper, we report the synthesis and detailed structural investigation of Na$_3$FeCr(PO$_4$)$_3$ (NFCP) sample containing three Na ions per formula unit. 
The room-temperature x-ray diffraction (XRD) reveals that the NFCP crystallizes in a monoclinic structure with long-range Na$^+$-vacancy ordering, consistent with other Na$_3$MM$'$(PO$_4$)$_3$ systems \cite{Park_CM_22, Chotard_CM_15}. Further, differential scanning calorimetry (DSC) identifies two reversible transitions: an $\alpha \Leftrightarrow \beta$ transition with multiple features between 333 and 353 K, followed by a $\beta \Leftrightarrow \gamma$ transition at 445 K ($\beta$- intermediate NASICON phase), corresponding to progressive disordering of the Na sublattice. Importantly, the [FeCr(PO$_4$)$_3$] framework remains essentially unchanged across these transitions, indicating that the structural evolution is governed by redistribution of Na ions rather than reconstruction of the polyanionic backbone \cite{Redhammer_ACB_21}. The transition can therefore be described as an order-disorder process within the Na conduction channels, where variations in Na-vacancy configurations modify the local electronic environment while preserving the overall framework. This coupling between ionic ordering, lattice symmetry, and local bonding highlights the central role of Na configurational degrees of freedom in determining phase stability and ion-transport behavior in NASICON-type polyanionic materials \cite{Wang_NC_23}.

\section{\noindent~Experimental}

The Na$_3$FeCr(PO$_4$)$_3$ sample is synthesized via the sol-gel method using stoichiometric amounts of NaH$_2$PO$_4$ ($\ge$99\%, Sigma-Aldrich), Fe(NO$_3$)$_3$·9H$_2$O ($\ge$98\%, Sigma-Aldrich), and Cr(NO$_3$)$_3$·9H$_2$O ($\ge$99\%, Alfa Aesar) as the precursors. To ensure a homogeneous precursor solution, the metal nitrates and phosphate salt were first dissolved in deionized water under continuous stirring at room temperature. The resulting solution was heated at 353 K under constant stirring to initiate gel formation. The obtained gel was dried overnight at 393 K and then finely ground using an agate mortar. The resulting powder was sintered at 973 K for 12 hrs in Argon environment to obtain the crystalline Na$_3$FeCr(PO$_4$)$_3$ phase.

The temperature dependent (303--523~K) synchrotron XRD measurements are performed at the BL-12 beamline of the Indus-2 synchrotron radiation facility, RRCAT, Indore using the incident X-ray wavelength of 0.76756~\AA. Powder synchrotron XRD data are analyzed by Rietveld refinement using the FullProf suite, and structural models are visualized through VESTA. The Raman spectroscopy measurements are performed at room temperature using a Renishaw inVia confocal microscope with a 532 nm excitation laser and a 2400 lines/mm grating. The X-ray photoelectron spectroscopy (XPS) measurements are performed on a Kratos AXIS SUPRA spectrometer using monochromatic Al K$\alpha$ radiation ($h\nu$ = 1486.6 eV). The DSC measurements are performed using a DSC2500 instrument (TA Instruments) by placing the samples in aluminum crucibles and then heating/cooling done at 10 K/min under nitrogen flow of 20 mL/min over the temperature range of 293--583~K. 

\section{\noindent~Results and discussions}

\begin{figure} 
\includegraphics[width=3.55in]{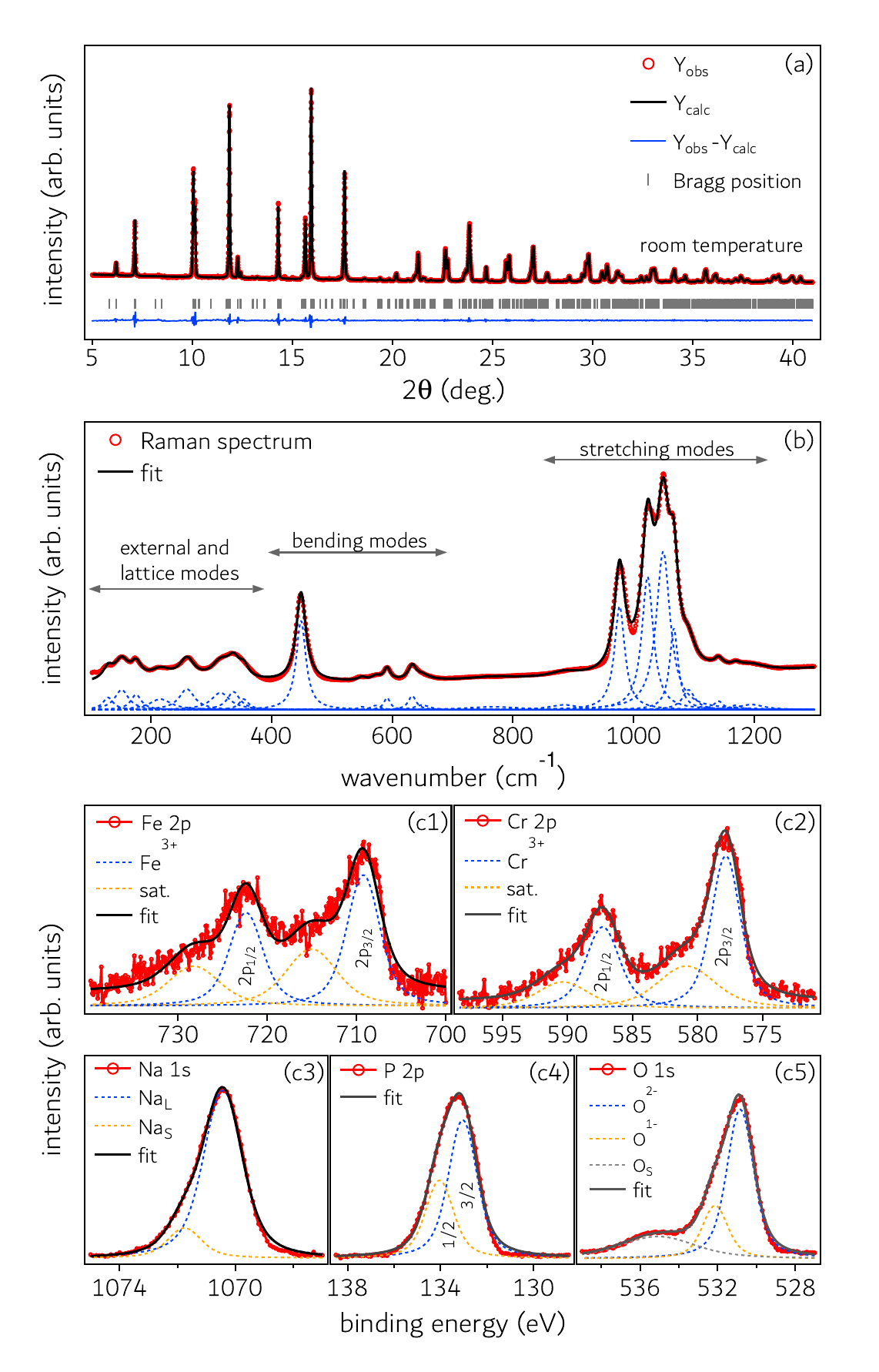}
\caption {(a) The Rietveld-refined synchrotron XRD pattern, (b) Raman spectrum, and (c1-c5) room-temperature core-level XPS spectra with deconvoluted components of Fe 2$p$, Cr 2$p$, Na 1$s$, P 2$p$, and O 1$s$ for the Na$_3$FeCr(PO$_4$)$_3$ sample.}
\label{Strc}
\end{figure}

The room-temperature synchrotron XRD pattern of Na$_3$FeCr(PO$_4$)$_3$ sample is shown in Fig.~\ref{Strc}(a). At first glance, the diffraction pattern closely resembles that of the rhombohedral NASICON structure with space group $R\bar{3}c$. However, closer inspection reveals the presence of an additional reflection at $\sim$6.2$^\circ$, which cannot be indexed within the rhombohedral symmetry. The appearance of this peak, together with several weak superstructure reflections, indicates a lowering of symmetry from the rhombohedral structure and can be indexed as the (-1~1~1) reflection of a monoclinic cell \cite{Masquelier_CM_00}. Such superstructure features are characteristic of Na$^+$/vacancy ordering and provide clear evidence for a monoclinic distortion in the NASICON framework. All the superstructure reflections can thus be successfully indexed in a monoclinic cell (space group $C2/c$) with lattice parameters $a=15.0657$~\AA, $b=8.6788$~\AA, $c=21.5565$~\AA, $\beta=90.159^\circ$, and $V/Z=234.88$~\AA$^3$, extracted through Rietveld refinement, as shown in Fig.~\ref{Strc}(a). Similar monoclinically distorted unit cells have been previously reported for several NASICON compounds, including 
Na$_3$FeV(PO$_4$)$_3$ \cite{Park_CM_21}. 

The Raman spectrum of Na$_3$FeCr(PO$_4$)$_3$ at room temperature is dominated by vibrations of the PO$_4^{3-}$ tetrahedra, as shown in Fig.~\ref{Strc}(b). The observed bands can be broadly classified into three regions: external and lattice modes below $\sim$400~cm$^{-1}$, PO$_4^{3-}$ bending modes in the 400--600~cm$^{-1}$ range, and strong P--O stretching vibrations between 900 and 1200~cm$^{-1}$ \cite{Singh_PRB_24} (assignments summarized in Table~S1 of \cite{SI}). The external modes arise from librational and translational motions of the PO$_4^{3-}$ groups, whereas the lattice modes involve the motion of Fe/Cr cations within their octahedral framework \cite{Barj_JSSC_92, Bih_JMS_09}. The internal tetrahedral vibrations are conventionally denoted as $\nu_1$, $\nu_2$, $\nu_3$, and $\nu_4$ \cite{Bih_JMS_09}. The $\nu_1$ mode at 976.8~cm$^{-1}$ and $\nu_3$ mode with peaks centered at 1049.1~cm$^{-1}$, correspond to symmetric and antisymmetric P--O stretching vibrations respectively, while the $\nu_2$ mode at 448.8~cm$^{-1}$ and $\nu_4$  peaks in range of 540-660 ~cm$^{-1}$ originate from symmetric and antisymmetric O--P--O bending motions, respectively \cite{Singh_PRB_24, Barj_JSSC_92}. The FE-SEM image, mass sum spectra, and the EDX imaging of the constituent elements are provided in Figs.~S1(a-g) of \cite{SI}. 


To elucidate the electronic structure of Na$_3$FeCr(PO$_4$)$_3$, core-level X-ray photoemission spectra of the constituent elements are presented in Figs.~\ref{Strc}(c1-c5). All the spectra are energy-calibrated relative to the C 1$s$ peak at 284.6 eV. Following the subtraction of the inelastic background using the Tougaard method, the Fe 2$p$, Cr 2$p$, Na 1$s$, P 2$p$, and O 1$s$ core-level spectra are fitted using Voigt line shapes (IGOR Pro software), representing a convolution of Gaussian and Lorentzian functions with a fixed shape factor of 0.6. The Fe 2$p$ core-level [Fig.~\ref{Strc}(c1)] exhibits well-resolved 2$p_{3/2}$ and 2$p_{1/2}$ main peaks at 712.7 and 725.9 eV, respectively, accompanied by satellite features at 718.6 and 731.8 eV. These spectral features are consistent with Fe present in the trivalent oxidation state \cite{Singh_PRB_24}. As shown in Fig.~\ref{Strc}(c2), the Cr 2$p$ core-level spectrum displays spin-orbit-split Cr 2$p_{3/2}$ and 2$p_{1/2}$ peaks at 577.8 and 587.3 eV, respectively, along with corresponding satellite features at 580.8 and 590.3 eV. These binding energies and satellite structures are characteristic of Cr in the 3+ oxidation state \cite{Biesinger_SIA_04, Sharma_ARXIV_26}. Fig.~\ref{Strc}(c3) presents the Na 1$s$ core-level, which is deconvoluted into two components located at 1070.4 and 1071.8 eV. While the strong peak (Na$_L$) at 1070.4 eV indicates the lattice sodium in the NASICON structure, the smaller signal (Na$_S$) likely originates from surface contributions \cite{Singh_PRB_24, Citrin_PRB_73}. The P 2$p$ core-level shown in Fig.~\ref{Strc}(c4) exhibits two characteristic peaks at 133.0 and 134.0 eV, corresponding to the 2$p_{3/2}$ and 2$p_{1/2}$ spin-orbit components, respectively. These binding energies are consistent with 5+ oxidation state of phosphorus within the tetrahedrally coordinated PO$_{4}^{3-}$ units \cite{Singh_PRB_24, Sharma_ARXIV_26}. Finally, the O 1$s$ core-level spectrum [Fig.~\ref{Strc}(c5)] is deconvoluted into three components. The dominant peak at 530.8 eV is attributed to lattice O$^{2-}$ species. The feature at 532.1 eV (O$^{1-}$) is assigned to oxygen vacancy-related species, while the higher-binding-energy component at 535.1 eV (O$_s$) is associated with surface hydroxyl (--OH) groups or adsorbed oxygen-containing species \cite{Kumar_JAP_20}. 

\begin{figure} 
\includegraphics[width=3.5in]{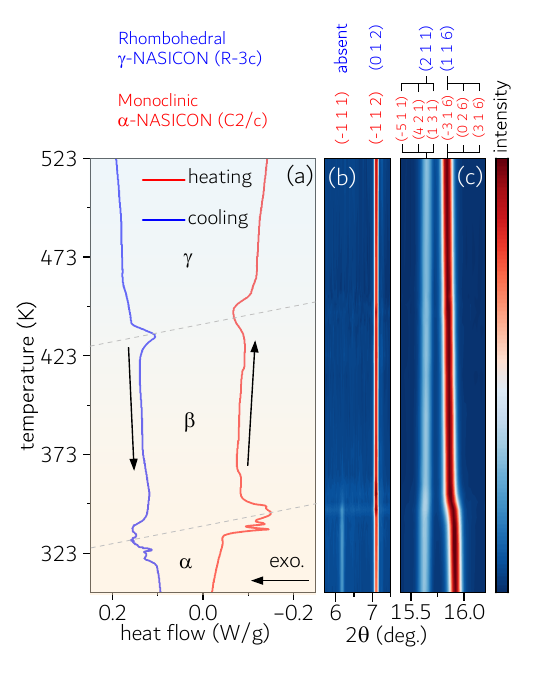}
\caption {(a) The differential scanning calorimetry (DSC) profiles of the Na$_3$FeCr(PO$_4$)$_3$ sample recorded during heating and cooling at a ramp rate of 10 K/min. (b, c) The contour plots of selected 2$\theta$ regions of {\it in-situ} synchrotron XRD (normalized) map showing the temperature-dependent evolution of diffraction intensities as a function of temperature.}
\label{DSC}
\end{figure}

Further, differential scanning calorimetry (DSC) is used to examine the thermal behavior of Na$_3$FeCr(PO$_4$)$_3$ sample. Fig.~\ref{DSC}(a) shows two reversible transitions observed during heating: a broad event centered around $\sim$350 consisting of several overlapping peaks, and a second transition near 445 K. Following earlier reports \cite{Kabbour_JACS_11, d’Yvoire_SSI_83, Chotard_CM_15, Park_CM_21}, the first one corresponds to the $\alpha$ (monoclinic) $\to$ $\beta$ (intermediate) transformation and is endothermic, whereas the $\beta \to \gamma$ (rhombohedral) transition is exothermic. The enthalpy change of the first transition (4.17 kJ mol$^{-1}$) is much larger than that of the second (-1.56 kJ mol$^{-1}$), indicating that most of the structural rearrangement occurs in the lower-temperature range. The changes in Na ion ordering govern these structural transformations: the monoclinic phase exhibits an ordered Na-vacancy arrangement, while the rhombohedral phase displays a largely disordered distribution \cite{Kabbour_JACS_11, Park_CM_21}. The complex multi-peak feature observed between 333 and 353 K suggests that the transformation does not occur in a single step. For comparison, in Na$_3$Fe$_2$(PO$_4$)$_3$ two well-separated endothermic transitions ($\alpha~\to~\beta$ at 368 K and $\beta~\to~\gamma$ at 418 K) are reported, whereas Na$_3$Cr$_2$(PO$_4$)$_3$ exhibits three sequential transitions ($\alpha~\to~\alpha^\prime~\to~\beta~\to~\gamma$) at 348, 411, and 439 K; though in both cases, no such multiple overlapping peaks were observed \cite{d’Yvoire_SSI_83}. In contrast, the Na$_3$V$_2$(PO$_4$)$_3$ exhibits, in addition to the prominent $\alpha \to \beta$ and $\beta^\prime \to \gamma$ transitions, an intermediate $\beta \to \beta^\prime$ transformation between 383 and 398 K \cite{Chotard_CM_15}. This transition is accompanied by splitting of DSC peaks, which have been attributed to the coexistence of composite structural domains during an incomplete transformation \cite{Chotard_CM_15}. In the present case, the overlapping peaks observed are likely originate from successive rearrangements of the Na sublattice, suggesting that the structural evolution proceeds through intermediate configurations than a single abrupt transition.

\begin{figure*} 
\includegraphics[width=7.1in]{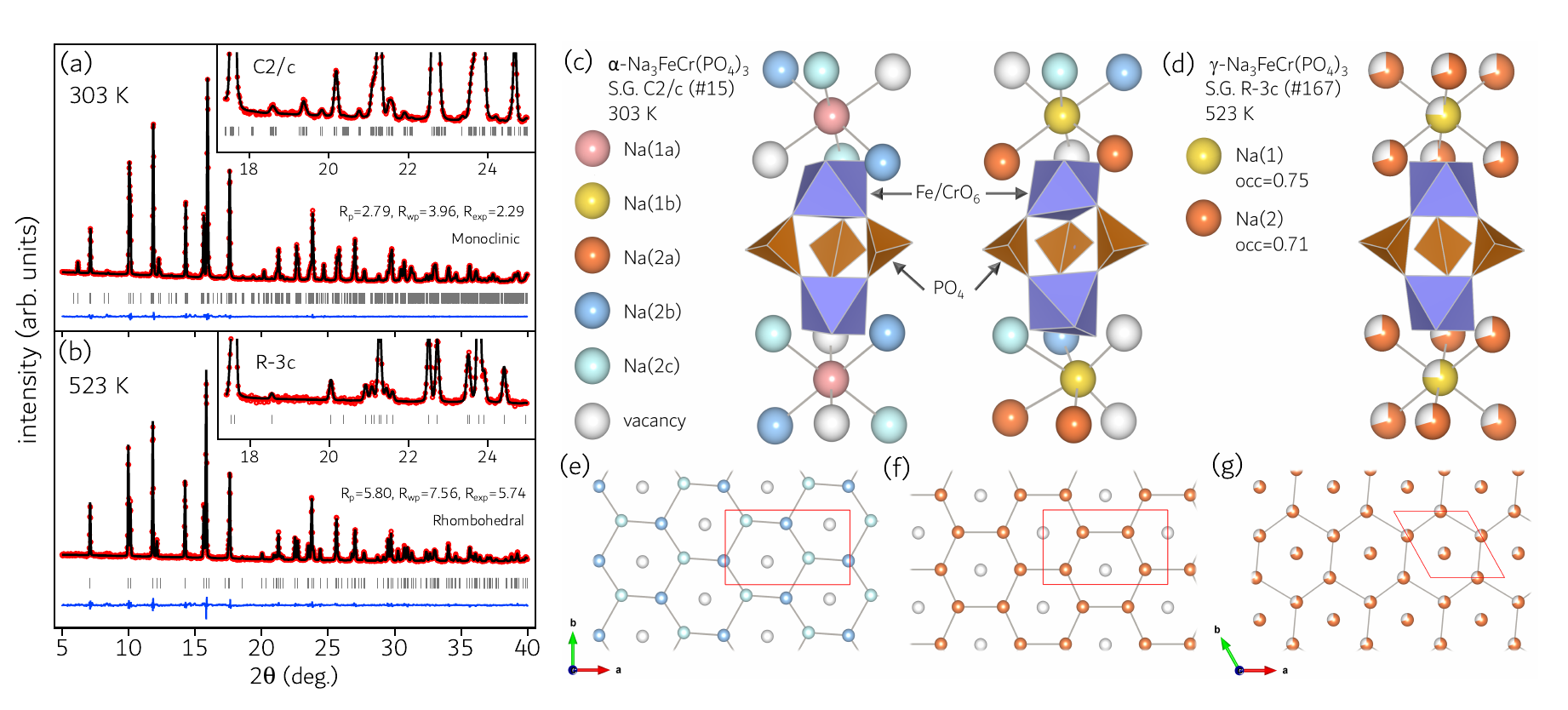}
\caption{The Rietveld refined synchrotron XRD patterns collected at (a) 303 K and fitted with the monoclinic $C2/c$ phase and (b) 523 K and fitted with the rhombohedral $R\bar{3}c$ phase. Depiction of sodium-site configurations in (c) $\alpha$-NFCP (the lantern-type structural units and the five unique Na positions together with the vacancy sites) and (d) $\gamma$-NFCP. Projection of the sodium sublattice in the ab plane showing ordered arrangement in (e) Na(2$b$)-Na(2$c$), (f) Na(2$a$) in $\alpha$-phase, and (g) statistically disordered Na(2) distribution in $\gamma$-phase. The red rectangle is the unit cell.}
\label{XRD}
\end{figure*}

In order to further understand the phase transition, {\it in-situ} temperature-dependent XRD measurements are performed in the range of 303--523~K. 
As clearly visible in Fig.~\ref{DSC}(b) where each pattern was normalized with respect to the maximum intensity, the superstructure reflection (-1 1 1), characteristic of the ordered monoclinic phase, appears at 6.2$\degree$ and completely disappears upon heating above 353 K, indicating the loss of long-range Na ordering. In addition, the unresolved monoclinic reflections (-3 1 6), (0 2 6), and (3 1 6) appearing at 15.9$\degree$ (at 303 K) represent symmetry-split components of the $R\bar{3}c$ (1 1 6) reflection, which are presented in Fig.~\ref{DSC}(c). Upon suppression of Na ordering and restoration of the higher-symmetry phase, these reflections, now represented by a single degenerate (1 1 6) peak, shift slightly towards lower 2$\theta$, and are consistent with the $R\bar{3}c$$\to$ $C2/c$ group-subgroup relationship. The following transformation matrix relates the $\gamma$ and $\alpha$ unit cell parameters \cite{Chotard_CM_15}.
\begin{equation}
\left(\begin{array}{ccc}
{a}_{C2/c} \\
{b}_{C2/c} \\
{c}_{C2/c} \\
\end{array}\right) = \left(
\begin{array}{ccc}
1 & -1 & 0 \\
1 & 1 & 0 \\
0 & 0 & 1 \\
\end{array}\right) 
\left(\begin{array}{ccc}
{a}_{R\bar{3}c} \\
{b}_{R\bar{3}c} \\
{c}_{R\bar{3}c} \\
\end{array}\right)
\label{transform}
\end{equation}
 or in a linear form, it can be written as:  
\begin{equation}
\begin{array}{ccc}
{a}_{C2/c} = {a}_{R\bar{3}c} - {b}_{R\bar{3}c} \\
{b}_{C2/c} = {a}_{R\bar{3}c} + {b}_{R\bar{3}c} \\
{c}_{C2/c} = {c}_{R\bar{3}c} \\
\end{array}
\end{equation}
%
%
The monoclinic C-centered superstructure contains 12 formula units per unit cell, whereas the rhombohedral phase contains 6 formula units. Despite this difference, the basic NASICON framework remains unchanged, composed of the same lantern units built from Fe/CrO$_6$ octahedra and PO$_4$ tetrahedra. The key effect of the reduced symmetry in the monoclinic superstructure is the increase in the number of crystallographically independent atomic positions. In the $\gamma$ phase, sodium occupies two sites, Na(1) and Na(2), corresponding to the 6$b$ and 18$e$ Wyckoff positions, respectively, giving 24 Na sites per unit cell. Upon symmetry lowering to the $\alpha$ phase, the number of sodium sites doubles to 48, with Na(1) splitting into Na(1$a$) (4$a$) and Na(1$b$) (8$f$) positions. Similarly, the Na(2) site splits into five independent Wyckoff positions, comprising one 4$e$ site and four 8$f$ sites. 

From the structural refinement of the monoclinic and rhombohedral phases, shown in Figs.~\ref{XRD}(a, b), the distribution of sodium ions within the lattice is determined. The monoclinic structure inherently contains three sodium ions per formula unit, of which one Na$^+$ fully occupies the Na(1) sublattice, distributed over the Na(1$a$) (4$a$) and Na(1$b$) (8$f$) sites. The remaining two sodium ions occupy the Na(2) sublattice, where three 8$f$ Wyckoff positions, labelled Na(2$a$), Na(2$b$), and Na(2$c$), are fully occupied. The other two Na(2) sites, Na(2$d$) and Na(2$e$), corresponding to the 8$f$ and 4$e$ Wyckoff positions, respectively, remain vacant. The corresponding atomic coordinates of the sodium sites, vacancies, and all the constituent atoms are provided in Table~\ref{tab:C2/c}. This ordered sodium distribution gives rise to inequivalent local environments around the Na(1$a$) and Na(1$b$) sites, and is visually depicted in Fig.~\ref{XRD}(c). Two distinct Na(1) coordination topologies coexist, corresponding to trans- and cis-type arrangements of the vacancies around crystallographically inequivalent Na(1$a$) (4$a$) and Na(1$b$) (8$f$) sites, respectively. Each Na(1) ion is coordinated by four fully occupied Na(2$a$, 2$b$, 2$c$) sites and two vacancies (Vac), forming distorted Na(1)-[Na(2)$_4$Vac$_2$] octahedra. In contrast, in the disordered $\gamma$ phase, see Fig.~\ref{XRD}(d), the Na(2) sites are statistically occupied, resulting in an averaged Na(1) coordination environment, similar to previous reports on the $\alpha$-NASICON phases \cite{Kabbour_JACS_11, Park_CM_21, Chotard_CM_15}. A projection of the sodium sublattice onto the $ab$ plane [Figs.~\ref{XRD}(e, f)] highlights the Na(2)-Vac arrangement more clearly and allows direct comparison with the $\gamma$ phase, see Fig.~\ref{XRD}(g). In both structures, the Na(2) cations form honeycomb-like layers in the $ab$ plane, stacked perpendicular to the $c$ axis. In the $\gamma$ phase, where Na ions are statistically distributed, the hexagonal units are centered by Na(2) cations. In contrast, in the $\alpha$ phase, the centers of these hexagons are occupied by vacancies, resulting in an ordered Na-Vac pattern. This redistribution breaks the threefold screw symmetry characteristic of the rhombohedral structure and stabilizes the lower-symmetry monoclinic phase. In addition to the in-plane ordering, three distinct Na(2)-Vac topological layers can be identified, as illustrated in Figs.~\ref{XRD}(e, f). Two layers are composed of alternating Na(2$b$)-Na(2$c$)-Vac arrangements, while the remaining one consists of Na(2$a$) sites interspersed with vacancies. Along the $c$ axis, six such Na(2)-Vac layers are stacked within a single unit cell in an $abca'b'c'$ sequence to describe the periodicity \cite{Kabbour_JACS_11}, as presented in Fig.~S2 of \cite{SI}. As reported by Chotard \textit{et al.} \cite{Chotard_CM_15}, each successive layer is shifted relative to the preceding one by approximately $(\tfrac{1}{6}a, \pm\tfrac{1}{6}b, \tfrac{1}{6}c)$ and simultaneously rotated by about 60$^\circ$ around the $c$ axis. This combined translation and rotation give rise to the three-dimensional Na-Vac ordering pattern observed in the monoclinic structure.

\begin{table*}[]
\centering
\caption{The refined structural parameters of Na$_3$FeCr(PO$_4$)$_3$ from the XRD patterns measured at 303 K and 523 K.}
\label{tab:C2/c}
\begin{tabular}{lp{2cm}p{2.6cm}p{2.6cm}p{2.6cm}p{2.5cm}p{1.2cm}}
\hline
	\multicolumn{7}{c}{Space group: $C2/c$ (\#15), Z=12, $a=15.0630$ \AA, $b=8.6778$ \AA, $c=21.5494$ \AA, $\beta=90.154^\circ$,} \\
	\multicolumn{7}{c}{$V=2816.79$ \AA$^3$, $V/Z=234.73$ \AA$^3$, $R_{p}=2.79\%$, $R_{wp}=3.96\%$, $R_{exp}=2.29\%$.}            \\ \hline
\text{Atom}   & \text{Wyckoff}   & \textbf{$x$}   & \text{$y$}  & \text{$z$}  & \text{$U$(\AA$^2$)}  & \text{Occupancy}  \\ \hline
Na(1a)     & 4a & 0.0000  & 0.0000  & 0.0000 & 0.045  & 1       \\
Na(1b)     & 8f & 0.3463  & 0.0201  & 0.1701 & 0.089  & 1       \\
Na(2a)     & 8f & 0.1836  & 0.1832  & 0.2573 & 0.037  & 1       \\
Na(2b)     & 8f & 0.0174  & 0.3034  & 0.4210 & 0.052  & 1       \\
Na(2c)     & 8f & 0.1761  & 0.1391  & 0.5846 & 0.007  & 1       \\
vacancy(a) & 8f & 0.2841  & 0.2748  & 0.0545 & 0      & 0       \\
vacancy(b) & 4e & 0.0000  & 0.3641  & 0.7500 & 0      & 0       \\
Fe/Cr(a)   & 8f & 0.0002 & -0.0021 & 0.3525 & 0.016 & 0.5/0.5 \\
Fe/Cr(b)   & 8f & 0.3324  & 0.0012  & 0.3141 & 0.000 & 0.5/0.5 \\
Fe/Cr(c)   & 8f & 0.3333  & 0.0022  & 0.0189 & 0.014 & 0.5/0.5 \\
P(a)       & 8f & 0.1855  & 0.1510  & 0.4168 & 0.002 & 1       \\
P(b)       & 4e & 0.0000  & 0.2917  & 0.2500 & 0.000 & 1       \\
P(c)       & 8f & 0.3488  & 0.3664  & 0.2540 & 0.016 & 1       \\
P(d)       & 8f & 0.5224  & 0.1435  & 0.0821 & 0.001 & 1       \\
P(e)       & 8f & 0.1691  & 0.2095  & 0.0852 & 0.011 & 1       \\
O(1a)      & 8f & 0.2628  & 0.4196  & 0.2464 & 0.007 & 1       \\
O(1b)      & 8f & 0.2507  & 0.1117  & 0.0726 & 0.005 & 1       \\
O(1c)      & 8f & 0.0883  & 0.4009  & 0.2486 & 0.008 & 1       \\
O(1d)      & 8f & 0.3523  & 0.1791  & 0.2581 & 0.000 & 1       \\
O(1e)      & 8f & 0.4297  & 0.0797  & 0.0809 & 0.004 & 1       \\
O(1f)      & 8f & 0.0820  & 0.1060  & 0.0845 & 0.000 & 1       \\
O(1g)      & 8f & 0.0956  & 0.0614  & 0.4122 & 0.019 & 1       \\
O(1h)      & 8f & 0.5056  & 0.3255  & 0.0938 & 0.000 & 1       \\
O(1i)      & 8f & 0.1777  & 0.3120  & 0.4149 & 0.031 & 1       \\
O(2a)      & 8f & 0.2424  & 0.1260  & 0.3557 & 0.000 & 1       \\
O(2b)      & 8f & 0.5807  & 0.1221  & 0.0242 & 0.000 & 1       \\
O(2c)      & 8f & 0.4060  & 0.3962  & 0.1988 & 0.000 & 1       \\
O(2d)      & 8f & 0.0112  & 0.1944  & 0.3122 & 0.050 & 1       \\
O(2e)      & 8f & 0.1518  & 0.3133  & 0.0373 & 0.058 & 1       \\
O(2f)      & 8f & 0.1829  & 0.3002  & 0.1469 & 0.021 & 1       \\
O(2g)      & 8f & 0.3995  & 0.4096  & 0.3105 & 0.033 & 1       \\
O(2h)      & 8f & 0.2273  & 0.0841  & 0.4776 & 0.024 & 1       \\
O(2i)      & 8f & 0.5655  & 0.0655  & 0.1334 & 0.015 & 1       \\ \hline
	\multicolumn{7}{c}{Space group: $R\bar{3}c$ (\#167), Z=6, $a=8.6847$ \AA, $c=21.7837$ \AA, $\gamma=120.00^\circ$,} \\
	\multicolumn{7}{c}{$V=1422.88$ \AA$^3$, $V/Z=237.15$ \AA$^3$, $R_{p}=5.80\%$, $R_{wp}=7.56\%$, $R_{exp}=5.74\%$.}            \\ \hline
\text{Atom}   & \text{Wyckoff}   & \text{$x$}   & \text{$y$}  & \text{$z$}  & \text{$U$(\AA$^2$)}  & \text{Occupancy}  \\ \hline
Na(1)     & 6b & 0.0000  & 0.0000  & 0.0000 & 0.103  & 0.75       \\
Na(2)     & 18e & 0.6325  & 0.0000  & 0.2500 & 0.043  & 0.71       \\
Fe/Cr     & 12c & 0.0000  & 0.0000  & 0.1481 & 0.000  & 0.5/0.5       \\
P          & 18e & 0.2901  & 0.0000  & 0.2500 & 0.000  & 1       \\
O(1)     & 36f & 0.1916  & 0.1672  & 0.0898 & 0.002  & 1       \\
O(2)     & 36f & 0.0321  & 0.2085  & 0.1934 & 0.012      & 1       \\ \hline
\end{tabular}%

\end{table*}

\begin{figure*} 
\includegraphics[width=7.0in]{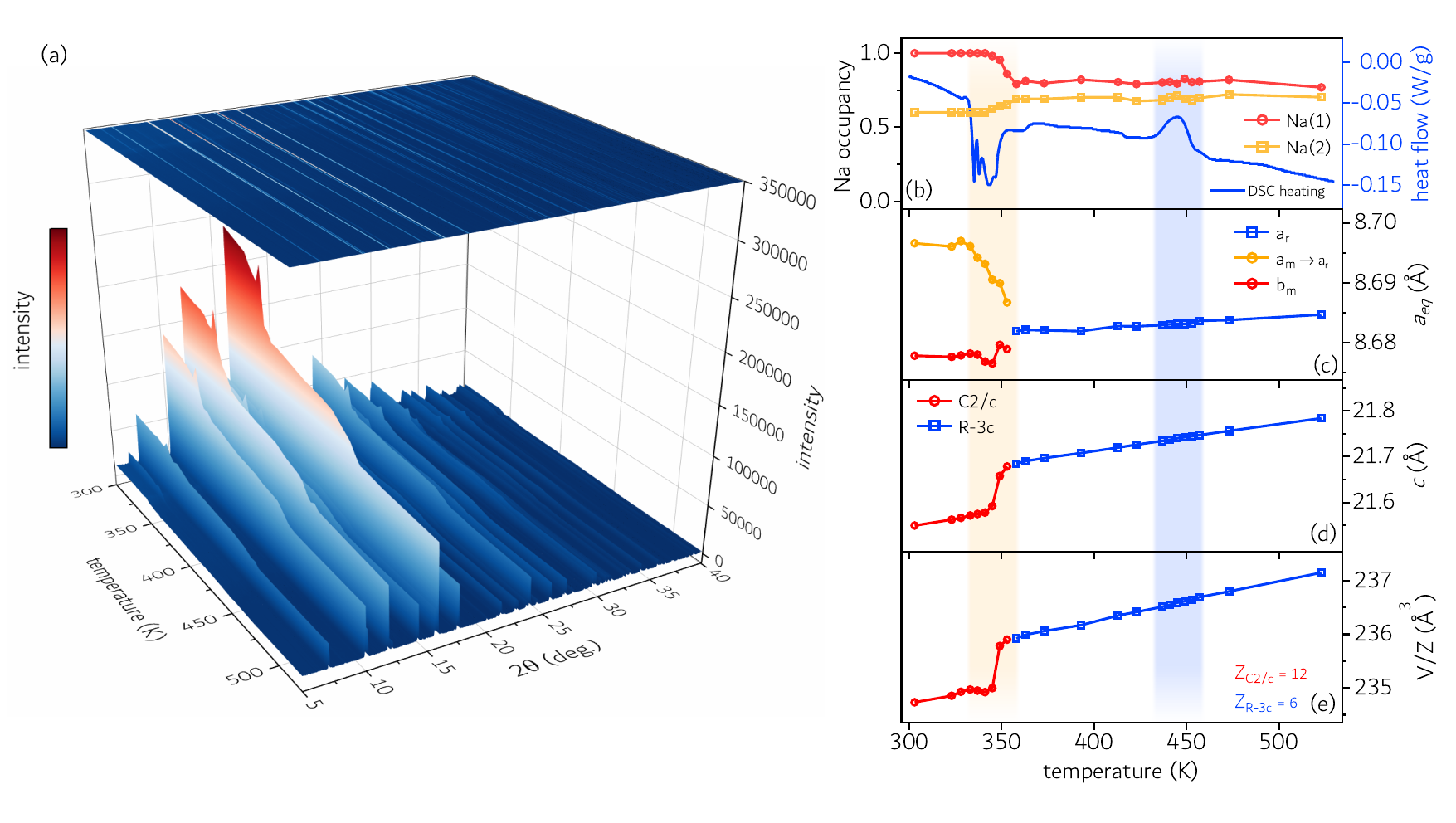}
\caption{(a) The three-dimensional map of intensity showing the evolution of diffraction peaks with temperature collected during heating from 303 to 523 K. (b) The temperature dependence of the site occupancy factor of Na(1) and Na(2) sites compared with the DSC heat-flow signal. (c--e) The evolution of lattice parameters across the transition, equivalent hexagonal $a_{eq}$, $c$, and unit-cell volume per formula unit (V/Z), respectively.}
\label{SXRD}
\end{figure*}

On the other hand, no detailed structural study of the intermediate $\beta$ phase has been reported till date \cite{Park_CM_21}. Notably, previous single-crystal studies have shown that a simple periodic structure cannot account for its diffraction pattern and instead requires multiple modulation vectors, indicating the coexistence of domains and an incomplete structural transformation \cite{Chotard_CM_15}. Therefore, for comparison, here the corresponding XRD patterns are refined using the $R\bar{3}c$ space group. The close similarity between the $\beta$ and $\gamma$ diffraction patterns suggests that their crystal structures are closely related, with only subtle changes in Na site occupancies expected across this transition. Moving forward, all the XRD patterns are analysed, and the fitted patterns are presented in Fig.~S3 of \cite{SI}. The three-dimensional temperature-resolved diffraction intensity map, presented in Fig.~\ref{SXRD}(a), provides a direct visualization of the structural evolution during heating. As discussed earlier, the gradual disappearance of the low-angle superstructure reflections ($\sim$6.2$\degree$) above $\sim$350~K indicates the loss of long-range Na-vacancy ordering associated with the monoclinic phase. Whereas, the fundamental reflections associated with the NASICON framework remain present throughout the entire temperature range, although a noticeable reduction in their overall intensity is observed near the transition. This behavior indicates that changes in diffraction intensity and the corresponding structural factor are strongly correlated with the redistribution of the Na sublattice, rather than by any significant modification of the polyanionic framework.  

\begin{table*}[]
\centering
\caption{The interatomic distances for Fe/Cr and Na in $\alpha$-Na$_3$FeV(PO$_4$)$_3$ and $\gamma$-Na$_3$FeV(PO$_4$)$_3$, along with the corresponding average bond lengths (Avg.), bond-valence sums (BVS), and polyhedral distortion parameter ($\Delta$).}
\label{tab:BVS}
\begin{tabular}{llllllllllll}
\hline
\multicolumn{9}{c}{$\alpha$-Na$_3$FeCr(PO$_4$)$_3$ (303K)}                                                 & \multicolumn{3}{c}{$\gamma$-Na$_3$FeCr(PO$_4$)$_3$ (523 K)} \\ \hline
Na(1a)       & O(1f) & 2 x 2.387 & Na(1b)       & O(1a) & 2.592 &              &       &       & Na(1)              & O(1)      & 6 $\times$ 2.510      \\
             & O(1g) & 2 $\times$ 2.441 &              & O(1b) & 2.665 &              &       &       &                    &           &                \\
             & O(1h) & 2 $\times$ 2.527 &              & O(1c) & 2.258 &              &       &       &                    &           &                \\
             &       &           &              & O(1d) & 2.348 &              &       &       &                    &           &                \\
             &       &           &              & O(1e) & 2.354 &              &       &       &                    &           &                \\
             &       &           &              & O(1i) & 2.595 &              &       &       &                    &           &                \\
Avg.         &       & 2.452     & Avg.         &       & 2.469 &              &       &       & Avg.               &           & 2.510          \\
BVS          &       & 1.05      & BVS          &       & 1.08  &              &       &       & BVS                &           & 0.888          \\
$\Delta$ (10$^{-4}$) &       & 5.58      & $\Delta$ (10$^{-4}$) &       & 38.77 &              &       &       & $\Delta$ (10$^{-4}$)       &           & 0              \\
Na(2a)       & O(1a) & 2.385     & Na(2b)       & O(1e) & 2.528 & Na(2c)       & O(1b) & 2.464 & Na(2)             & O(1)      & 2 $\times$ 2.362      \\
             & O(1a) & 2.427     &              & O(1f) & 2.278 &              & O(1f) & 2.555 &                    & O(1)      & 2 $\times$ 2.442      \\
             & O(1c) & 2.380     &              & O(1g) & 2.417 &              & O(1h) & 2.595 &                    & O(2)      & 2 $\times$ 2.628      \\
             & O(1d) & 2.541     &              & O(1i) & 2.420 &              & O(1i) & 2.244 &                    & O(2)      & 2 $\times$ 2.807      \\
             & O(2a) & 2.350     &              & O(2b) & 2.502 &              & O(2a) & 2.703 &                    &           &                \\
             & O(2c) & 2.988     &              & O(2d) & 2.530 &              & O(2b) & 2.837 &                    &           &                \\
             & O(2d) & 2.858     &              & O(2e) & 2.705 &              & O(2g) & 2.568 &                    &           &                \\
             & O(2f) & 2.586     &              & O(2i) & 2.846 &              & O(2h) & 2.480 &                    &           &                \\
Avg.         &       & 2.564     & Avg.         &       & 2.528 & Avg.         &       & 2.556 & Avg.               &           & 2.560          \\
BVS          &       & 1.19      & BVS          &       & 1.24  & BVS          &       & 1.16  & BVS                &           & 1.145          \\
$\Delta$ (10$^{-4}$) &       & 75.56     & $\Delta$ (10$^{-4}$) &       & 42.87 & $\Delta$ (10$^{-4}$) &       & 40.96 & $\Delta$ (10$^{-4}$)       &           & 45.35          \\
Fe/Cr(1a)    & O(1f) & 2.061     & Fe/Cr(1b)    & O(1a) & 2.060 & Fe/Cr(1c)    & O(1b) & 1.949 & Fe/Cr              & O(1)      & 3 $\times$ 2.016      \\
             & O(1g) & 2.003     &              & O(1c) & 2.006 &              & O(1e) & 2.082 &                    & O(2)      & 3 $\times$ 1.957      \\
             & O(1h) & 1.892     &              & O(1d) & 1.983 &              & O(1i) & 2.189 &                    &           &                \\
             & O(2c) & 2.002     &              & O(2a) & 1.954 &              & O(2b) & 1.926 &                    &           &                \\
             & O(2d) & 1.922     &              & O(2f) & 1.950 &              & O(2e) & 2.020 &                    &           &                \\
             & O(2g) & 1.924     &              & O(2i) & 1.987 &              & O(2h) & 1.974 &                    &           &                \\
Avg.         &       & 1.967     & Avg.         &       & 1.990 & Avg.         &       & 2.023 & Avg.               &           & 1.987          \\
BVS &
   &
  \begin{tabular}[c]{@{}l@{}}3.46/3.15\\ (Fe/Cr)\end{tabular} &
  BVS &
   &
  \begin{tabular}[c]{@{}l@{}}3.23/2.94\\ (Fe/Cr)\end{tabular} &
  BVS &
   &
  \begin{tabular}[c]{@{}l@{}}3.02/2.75\\ (Fe/Cr)\end{tabular} &
  BVS &
   &
  \begin{tabular}[c]{@{}l@{}}3.26/2.96\\ (Fe/Cr)\end{tabular} \\
$\Delta$ (10$^{-4}$) &       & 8.96      & $\Delta$ (10$^{-4}$) &       & 3.44  & $\Delta$ (10$^{-4}$) &       & 19.69 & $\Delta$ (10$^{-4}$)       &           & 2.26           \\ \hline
\multicolumn{12}{c}{Bond valence sum (BVS) is calculated from the Zachariasen formula, equation S1 of \cite{SI},} \\
\multicolumn{12}{c}{and polyhedral distortion ($\Delta$) is calculated using equation S2 of \cite{SI}}  \\ \hline
\end{tabular}
\end{table*}

We note that near a structural transition, diffraction intensities primarily reflect the relative fraction of coexisting phases. Since the two structures are related through a group-subgroup symmetry relation, the lower-symmetry space group can consistently describe both states and therefore provides a reliable basis for structural refinement across the transition. During the analysis near the transition, the XRD patterns are refined using the monoclinic phase. Subsequently, the sodium ions were introduced into the vacant sites, namely Na(2$d$) and Na(2$e$), and the structure was refined again. The site occupancy factor (SOF) of Na(2$d$) and Na(2$e$) collapses to nearly zero for 337 and 341 K, while the SOF for other sodium sites remains unchanged. At the 345 K, the sodium ion starts to fill the Na(2$d$) and Na(2$e$) sites, with the SOF values of 0.17 and 0.11, respectively, and as the temperature is increased, the SOF values continue to increase. The average Na(1) and Na(2) SOF for monoclinic phase are calculated by using equations below,  
\begin{equation}
Na(1)_{avg} = \frac{4\times Na(1a)+8\times Na(1b)}{12}
\end{equation}
\begin{equation}
Na(2)_{avg} = \frac{\left[8\times \sum_{i}^{a,b,c,d}Na(2i)\right]+4\times Na(2e)}{36}
\end{equation}
which are found to be same for the rhombohedral phase. The average site-occupation factors (SOF) of the Na(1) and Na(2) sites are shown in Fig.~\ref{SXRD}(b), together with the DSC curve recorded during the heating. Notably, the sodium rearrangement across the transition involves not only redistribution between Na(2) sites and vacancies, but also migration of Na ions from the Na(1) sites to partially fill the Na(2) sublattice. Consequently, the average occupancy of Na(2) increases at the expense of the Na(1) sites. In the NASICON frameworks, the Na(1) sites having 6 oxygen coordination are generally considered the most energetically favorable positions and are typically occupied preferentially before other Na sites. This behavior is commonly observed in NASICON compounds without the Na ordering \cite{Wang_JMCA_22}. However, in ordered NASICON systems, several studies have reported partial occupancies of both Na(1) and Na(2) sites in the high-temperature disordered $\gamma$ phase \cite{Park_CM_21, Chotard_CM_15}, where the reduced Na(1) occupancy is attributed to the transfer of Na ions into the Na(2) sites. The refined occupation factors in the disordered $\gamma$ phase (at 523 K) are 0.75 and 0.71 for the Na(1) and Na(2) sites, respectively, whereas in the ordered $\alpha$ phase (at 303 K), the Na(1) site is found to be fully occupied. To further examine the local bonding environment of the sodium sites, the bond lengths, bond-valence sums (BVS), and polyhedral distortion parameter ($\Delta$) are summarized in Table~\ref{tab:BVS}. In the $\alpha$ phase, the average Na(1$a$)--O and Na(1$b$)--O bond lengths are 2.452 and 2.469~\AA, with corresponding BVS values of 1.05 and 1.08, respectively. In contrast, the $\gamma$ phase exhibits a larger average Na(1)--O bond length of 2.510~\AA\ and a reduced BVS value of 0.888, indicating that the Na ions are more weakly bound at the Na(1) site in the high-temperature phase. By comparison, the Na(2) sites show only minor variations in both the average bond lengths and their corresponding BVS values between the two phases. The weakening of the Na(1) coordination environment in the $\gamma$ phase, therefore, reflects the increasing configurational freedom of the Na sublattice at elevated temperatures and a greater tendency towards mobility \cite{Masquelier_CM_00}. These observations are consistent with the theoretical analysis reported by Wang \textit{et al.} \cite{Wang_JMCA_22}, which show that a larger configurational entropy can be achieved when Na ions are distributed over both the Na(1) and Na(2) sites rather than being confined only to the Na(2) sublattice. On the other hand, when the Na(1) sites are fully occupied, their contribution to the configurational entropy becomes negligible, and the overall entropy arises solely from the distribution of Na ions within the Na(2) sites.  

To further examine how this redistribution of Na ions influences the overall crystal structure, we plot the temperature evolution of the lattice parameters and unit-cell volume in Figs.~\ref{SXRD}(c--e). For a direct comparison across the symmetry change, the monoclinic lattice parameter $a_m$ is transformed to the equivalent hexagonal rhombohedral setting $a_r^m$ using the transformation matrix given in equation~\ref{transform}. The transformed $a_r^m$ parameter, together with the monoclinic $b_m$ and the rhombohedral $a_r$, are plotted as a function of temperature in Fig.~\ref{SXRD}(c). A monotonic decrease in the transformed lattice parameter $a_r^m$ is observed as the temperature approaches the first transition (around 350~K), until it nearly converges with the rhombohedral $a_r$ value, after which it remains essentially constant with further heating. In contrast, the $c$-axis exhibits a pronounced discontinuity near 350~K, followed by a continuous thermal expansion over the remaining temperature range as depicted in Fig.~\ref{SXRD}(d). Although a slight compression is observed within the $ab$ plane relative to the expansion along the $c$ axis, the gradual change in $a_r^m$ begins at lower temperatures and precedes the abrupt increase in the $c$ parameter, indicating anisotropic thermal expansion of the unit cell \cite{Kumar_JPCL_22}. The increase in the $c$ dimension during the transformation can be primarily attributed to the enlargement of the Na(1) cavity height, as the Na(1) site becomes progressively depopulated with increasing temperature, leading to enhanced O--O repulsion within the Na(1) cavity along this direction, while the average Fe/Cr--O bond lengths exhibit only minor changes (see Table~\ref{tab:BVS}) \cite{Park_CM_22, Masquelier_CM_00}. The average Fe/Cr--Fe/Cr distance along the $c$ axis within a single lantern unit is 4.416 and 4.439~\AA\ in the $\alpha$ and $\gamma$ phases observed at 303~K and 523~K, respectively. However, when considering two adjacent lantern units along the $c$ axis separated by the Na(1)O$_6$ cavity, the average Fe/Cr--Fe/Cr distance increases from 6.359~\AA\ in the $\alpha$ phase to 6.453~\AA\ in the $\gamma$ phase. This change indicates that the expansion along the $c$ direction is primarily associated with the enlargement of the Na(1) cavity rather than significant expansion of the lantern unit. Consistent with this observation, the octahedral distortion of the Fe/CrO$_6$ units is substantially reduced in the $\gamma$ phase, with a distortion parameter $\Delta =$ 2.26, compared to the much larger average distortion ($\Delta=$ 10.7) in the $\alpha$ phase, which contributes to the monoclinic distortion. The evolution of the unit-cell volume normalized by the number of formula units (V/Z) as a function of temperature [see Fig.~\ref{SXRD}(e)] indicates a clear discontinuity across the first transition ($\alpha \Leftrightarrow \beta$), in contrast, only gradual and continuous changes in the lattice parameters and V/Z are observed near the second thermal anomaly at $\sim$445 K ($\beta \Leftrightarrow \gamma$). Moreover, no significant variation in the refined Na-site occupancies is detected in this temperature range. Despite the absence of a clear crystallographic signature in the diffraction data, the DSC measurements reveal a distinct thermal signal, accompanied by a small reduction in the overall diffraction intensity. This suggests that the second anomaly may not be associated with a symmetry change of the average crystal structure but rather reflects a subtle rearrangement or short-range configurational fluctuations that do not alter the long-range periodic structure detectable by Bragg diffraction.

\begin{figure} 
\includegraphics[width=3.4in]{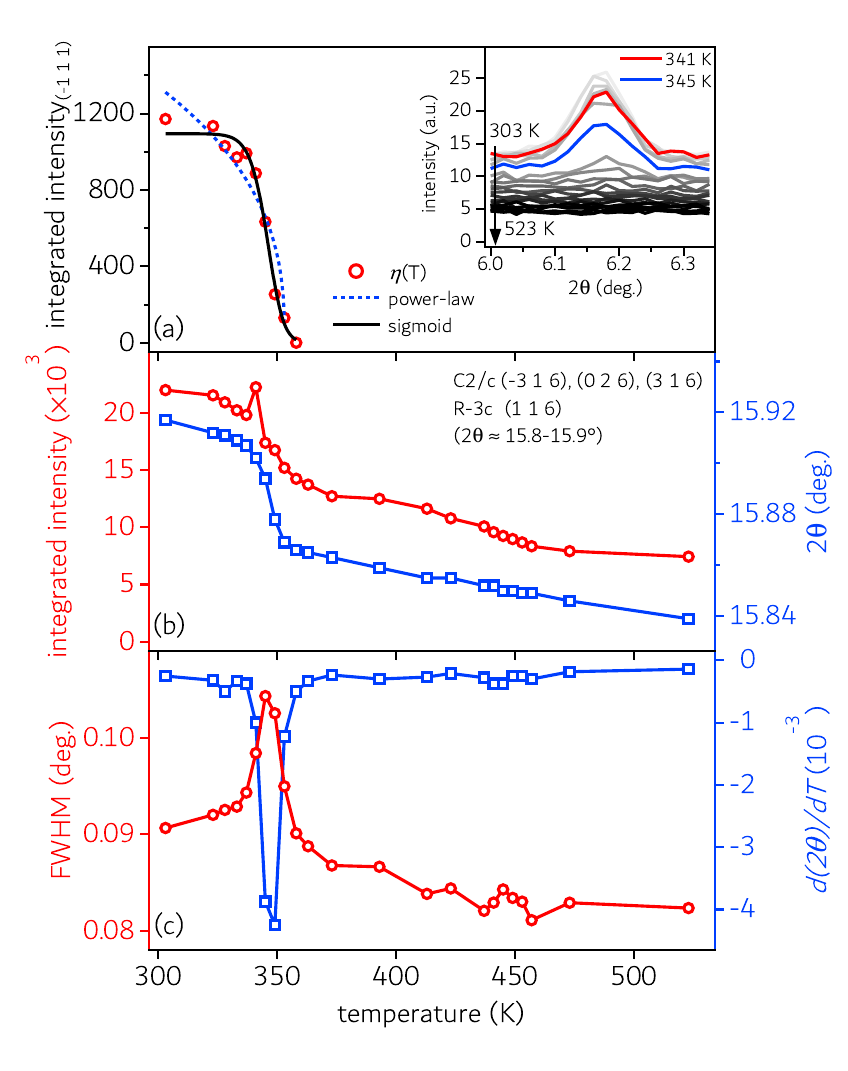}
\caption{(a) The temperature evolution of the integrated intensity of the superstructure reflection (-1~1~1) evidencing the loss of Na ordering across the transition, and inset showing the enlarged view of peak profile highlighting changes across the transition region (341-345 K). (b) The temperature dependence of the integrated intensity and peak position (2$\theta$) of the strongest Bragg reflections [(-3 1 6), (0 2 6), (3 1 6) for $C2/c$ and (1 1 6) for $R\bar{3}c$]. (c) The corresponding full width at half maximum (FWHM) and temperature derivative of 2$\theta$, highlighting the anomaly at the structural phase transition.}
\label{Fig5}
\end{figure}

So far, the analysis indicates that the observed structural transformation is primarily driven by the redistribution of the Na sublattice, while the polyanionic framework remains essentially unchanged. Moreover, the most significant rearrangements within the Na sublattice occur during the first transition. Therefore, the following discussion will focus mainly on the temperature range associated with the first $\alpha \Leftrightarrow \beta$ transition. For an order-disorder structural transition, the order parameter ($\eta$) can be represented by the integrated intensity of a superstructure reflection ($I\propto\eta^2$), which in the present case corresponds to the (-1~1~1) peak. The temperature dependence of this intensity is shown in Fig.~\ref{Fig5}(a), where a rapid collapse is observed near the transition; the corresponding diffraction profiles are provided in the inset. For a continuous order-disorder transition, the temperature dependence of the order parameter is commonly described within the Bragg-Williams mean-field framework \cite{Bragg_MPC_34, Subbaswamy_PRB_78}, which yields the self-consistent relation 
\begin{equation}
\eta = tanh\left(\frac{zJ}{k_BT}\eta\right)
\end{equation}
where $z$ is the number of interacting nearest neighbor sites, $J$ is the interaction energy between the Na sites, and $k_B$ is the Boltzmann constant. Here, linearization near the transition ($\eta~\to~0$) yields the critical condition $zJ/k_BT_c = 1$ giving $T_c = zJ/k_B$. Thus, the transition temperature reflects the strength of configurational interactions within the Na sublattice. Furthermore, close to the transition temperature ($T \lesssim T_c$), expansion of the hyperbolic tangent yields the mean-field critical behavior 
\begin{equation}
\eta^2 \propto (T_c-T)
\end{equation}
Since the integrated intensity of the superstructure reflection is proportional to the square of the order parameter, mean-field theory predicts a linear scaling $I\propto(T_c - T)$ close to the transition temperature. To assess whether the present transformation follows continuous critical behavior, the temperature dependence of the superstructure intensity has been analyzed using a power-law expression of the form $I\propto(T_c - T)^{2\beta}$ for $T < T_c$ \cite{Veiga_PRB_20, Mishra_PRB_11}. However, the experimental data deviate systematically from this critical scaling, see Fig.~\ref{Fig5}(a), and no stable convergence is achieved. This behavior is consistent with the DSC results, which display multiple thermal anomalies, indicating that the transformation proceeds through several intermediate steps. It is important to note that powder XRD probes the spatially averaged periodic structure over a macroscopic volume of the sample, and near the transition, the coexisting phases will reflect the macroscopic picture in the XRD patterns and not the microscopic ordering. This suggests the temperature dependence of the measured order parameter will present the average phase fraction and can be written as $\eta(T) = f(T)\cdot\eta_o$, where $f(T)$ represents the fraction of the ordered phase and $\eta_o$ is the order parameter of the fully ordered state. Since the structural framework remains essentially unchanged across the transition and only the Na sublattice undergoes configurational rearrangement, the evolution of $f(T)$ is governed by the redistribution of Na ions between energetically favorable ordered sites and statistically equivalent disordered positions. Within this picture, the Na sites may be either occupied or vacant, forming a binary configurational system. The probability of site occupation is therefore determined by the balance between interaction energy and thermal fluctuations. In the mean-field limit, this occupation probability follows a Fermi-Dirac type distribution \cite{Khachaturyan_DP_13}, and the order parameter becomes 
\begin{equation}
\eta(T) = \frac{\eta_o}{1+exp\left(\frac{-\mu}{k_BT}\right)}
\end{equation}
where, $\mu$ is the chemical potential and $k_B$is the Boltzmann constant. For simplification, we present the $\mu$ as the change in the Gibbs free energy ($\Delta G$) during this order-disorder transformation \cite{Chen_MRSB_19}. Expanding $\Delta G$ about the transition temperature yields $\Delta G=-\frac{L}{T_c}(T-T_c)$; $\Delta G$=0 at $T=T_c$, where L is the latent heat. Substituting $\mu$ with the obtained expression for $\Delta G$, the order parameter near $T_c$ becomes 
\begin{equation}
\eta(T) = \frac{\eta_o}{1+exp\left[\frac{L}{k_BT_c}\left(\frac{T-T_c}{T}\right)\right]}
\end{equation} 
and can be written as 
\begin{equation}
\eta(T) = \frac{\eta_o}{1+exp\left[\alpha\left(\frac{T-T_c}{T}\right)\right]}
\end{equation} 
The dimensionless parameter $\alpha$ = L/($k_BT_c$) represents the ratio between the ordering enthalpy and thermal energy at the transition temperature. The order parameter is fitted using the above expression, which gave a transition temperature $T_c$ = 349 K, and $\alpha$ = 0.0121. 

Moreover, the evolution of the reflections in the 2$\theta \approx 15.8$--15.9$^\circ$ region is shown in Fig.~\ref{Fig5}(b). As discussed above, this region contains the unresolved monoclinic $C2/c$ reflections (-3~1~6), (0~2~6), and (3~1~6), which transform into the rhombohedral $R\bar{3}c$ (1~1~6) reflection upon heating. As the temperature approaches the first transition, the integrated intensity decreases abruptly, accompanied by a shift in the peak position, reflecting the structural rearrangement associated with the $\alpha \rightarrow \beta$ transformation. But, around the second transition, only a gradual decrease in the integrated intensity is observed, without any significant change in the peak position, indicating that this transition does not involve a substantial modification of the average lattice. Although a detailed description of this effect is not readily available in the literature, our XRD analysis suggests that the structural factors depend not only on the temperature-induced lattice vibrations (Debye-Waller factor), but also on the degree of localization of the Na ions \cite{Sharma_PRB_21}. In the ordered monoclinic structure, the Na ions occupy well-defined Na(1) and Na(2) sites, whereas in the high-temperature disordered state, they are statistically distributed over the available sites. This redistribution modifies the corresponding structural factors and contributes to the observed changes in diffraction intensities. However, a more rigorous analysis would be required to quantitatively establish this correlation. 
Further insight into the transition mechanism is obtained from the temperature evolution of the peak width and position shown in Fig.~\ref{Fig5}(c). The FWHM [$C2/c$ (-3~1~6), (0~2~6), and (3~1~6), $R\bar{3}c$ (1~1~6) reflections] exhibits a pronounced maximum near the transition, indicating enhanced lattice strain and domain coexistence associated with the redistribution of Na ions. Also, the derivative of the peak position with respect to temperature, $d(2\theta)/dT$, displays a sharp anomaly at nearly the same temperature, signaling an abrupt change in the lattice response. It is well established that uniform stress primarily shifts diffraction peak positions, whereas non-uniform strain leads to peak broadening \cite{Dolabella_SM_22, Abe_JPCM_03}. The peak fitting yields transition temperatures of 346 and 347 K from the maxima in FWHM and $d(2\theta)/dT$ curves, respectively. These observations indicate that the transition proceeds through a coexistence regime involving short-range ordering and nanoscale domains that generate nonuniform strain, followed by the growth of the high-temperature phase, which shifts the peak position. Notably, other reflections such as (-1~1~2) in $C2/c$ [(0~1~2) in $R\bar{3}c$, see Fig.~\ref{DSC}(b)] show negligible peak shift, suggesting anisotropic lattice stress, consistent with the larger change observed along the $c$ axis compared to the $a$ and $b$ directions. The close agreement between these structural indicators and the transition temperature ($T_c = 349$ K) obtained from the sigmoidal fitting of the order parameter, and the clear correlation between temperature dependent Na-vacancy ordering, lattice strain, and symmetry lowering, confirms that the transformation is driven by the rearrangement of the Na sublattice.

\section{\noindent~Conclusions}

In summary, we investigated the temperature-driven structural evolution of Na$_3$FeCr(PO$_4$)$_3$ using synchrotron based X-ray diffraction measurements. At room temperature, Na$_3$FeCr(PO$_4$)$_3$ crystallizes in the monoclinic space group $C2/c$ ($a=15.0657$~\AA, $b=8.6788$~\AA, $c=21.5565$~\AA, $\beta=90.159^\circ$, $Z=12$) 
and exhibits superstructure reflections arising from long-range Na$^+$ ordering. Upon heating, the structure transforms to a rhombohedral $R\bar{3}c$ phase with statistically disordered Na ions, while the [FeCr(PO$_4$)$_3$] framework remains essentially unchanged, indicating that the transition is governed by redistribution of the Na sublattice rather than reconstruction of the polyanionic network. This transformation is accompanied by a progressive transfer of Na ions from the six-coordinated Na(1) sites to the eight-coordinated Na(2) sites due to the weakening of Na(1) coordination environment in the high-temperature $\gamma$ phase. The integrated intensity of the superstructure reflection (-1~1~1), which is proportional to the square of the order parameter, decreases rapidly between 333 and 353 K and vanishes near $\sim$358 K. The collapse of the superstructure intensity near the transition is accompanied by a discontinuous increase in the $c$-axis lattice parameter and unit-cell volume. The order parameter does not follow mean-field critical scaling; instead, calorimetric measurements reveal multiple overlapping thermal anomalies, indicating a finite coexistence regime involving intermediate Na sublattice configurations. The temperature dependence of the superstructure intensity is well described by a sigmoidal phase-fraction model, yielding a characteristic transition temperature of $T_c \approx 349$~K. These results demonstrate the strong coupling between the configurational entropy of Na ions and symmetry lowering in NASICON frameworks. 

\section{\noindent~Acknowledgements}

MS acknowledges CSIR-HRDG for fellowship support. RSD acknowledges the Department of Science and Technology (DST) for financial support through a core research grant (file no. CRG/2020/003436). The authors acknowledge IIT Delhi for providing access to XPS, FESEM, and DSC facilities at the Central Research Facility (CRF), as well as Raman spectroscopy facilities at the Department of Physics. We also acknowledge RRCAT, Indore for temperature-dependent XRD measurements. 

\end{document}